\documentclass[a4paper,12pt]{article}
\usepackage{natbib}
\usepackage{graphicx}	
\usepackage{amsmath}	
\usepackage{amssymb}	
\usepackage{textcomp}
\usepackage{color}

\RequirePackage[colorlinks=true
,urlcolor=blue
,anchorcolor=blue
,citecolor=blue
,filecolor=blue
,linkcolor=blue
,menucolor=blue
,pagecolor=blue
,linktocpage=true
,pdfproducer=medialab
,pdfa=true
]{hyperref}
\usepackage{mwe}
\usepackage{array}
\usepackage[T1]{fontenc} 
\usepackage{mathtools,euscript}  

\title{An Accurate Reconstruction of 
CMB E Mode Signal over Large Angular Scales using Prior Information of  CMB  
Covariance Matrix in ILC Algorithm}
\author{Ujjal Purkayastha$^1$\footnote{email:ujjalp@iiserb.ac.in}, Vipin Sudevan$^1$\footnote{email:vipins@iiserb.ac.in} and Rajib Saha$^1$\footnote{email:rajib@iiserb.ac.in} }
\date{}
\begin{document}
\maketitle
\begin{center}
{$^{1}$Department of Physics, Indian Institute of Science Education and Research, Bhopal-462066, India\\ }
\end{center}

\begin{abstract}

In the recent years, the internal-linear-combination (ILC) method was investigated 
extensively in the context of reconstruction  of Cosmic Microwave
Background (CMB) temperature anisotropy signal using  observations obtained  by WMAP and Planck satellite
missions. In this article, we, for the first time,
apply the ILC  method to reconstruct the large scale CMB E mode polarization signal, which serves as the 
unique probe of ionization history of the Universe, using simulated observations of  $15$ frequency
CMB polarization maps of future generation Cosmic Origin Explorer (COrE) satellite mission. We find that 
usual ILC cleaned E mode map is highly erroneous due to presence of a chance-correlation between CMB 
and astrophysical foreground components in the 
empirical covariance matrix which is used to estimate the weight factors. The cleaned angular
power spectrum for E mode is strongly biased and erroneous  due to these chance correlation factors.  In order
to address the issues of bias and errors we extend and improve the usual ILC method for CMB E mode reconstruction by  incorporating
prior information of theoretical E mode angular power spectrum while estimating the weights 
for linear combination of input maps (Sudevan \& Saha 2018b). 
Using the E mode covariance matrix 
effectively suppresses the CMB-foreground chance correlation power leading   to  
an accurate reconstruction of cleaned CMB E mode map and its angular power spectrum. We provide a comparative 
study of the performance of the usual ILC and the new method over  large 
angular scales of the sky and show that the later produces significantly statistically 
improved results  than the former.  The new E mode CMB angular power spectrum contains
neither any significant negative bias at the low multipoles nor any  positive 
foreground bias at relatively higher mutlipoles. The error estimates of the cleaned spectrum agree 
very well with the cosmic variance induced error.

\end{abstract}


\section{Introduction}

Weakly polarized  component of the Cosmic Microwave Background (CMB) anisotropy ~\citep{Rees,Basko}
serves as an important probe~\citep{sp_zal1997,zal_sel1998,knox2002} to understand different 
evolutionary epochs in the expanding Universe. The E mode component of the polarized CMB signal over large 
angular scales of the sky is sensitive to interactions of CMB photons with 
the free electrons in space and therefore is a potential observable to 
tightly constrain the ionization history of the Universe~\citep{page2007,Burigana2008, planck_pol2016,Watts2019}. On the very 
large angular scales the CMB E mode signal is dominated by the secondary anisotropies 
generated during the scattering of CMB photons with the high energy electrons produced 
after galaxies were formed. The peak height and location of EE mode CMB angular
power spectrum on the large angular scales are determined by the amount of the free electrons 
present in the inter galactic medium and epoch of reionization respectively. On the scales $\ell \lesssim 15$
the EE mode spectrum can be used to constrain the physics at the last scattering 
surface as well. In the era of precision measurement of cosmological parameters 
CMB E mode signal provide accurate measures of amplitude of scalar field 
fluctuations and the optical depth of reionization epoch by breaking degeneracies between the 
two.

In order to extract the complete set of cosmological information available 
from CMB E mode signal on large angular scales one needs to perform 
an accurate reconstruction of the CMB  E mode map on these scales. This, however, 
is a challenging task since the weak E mode CMB signal remains hidden inside 
the weakly understood, strong polarized foreground signals due to synchrotron and thermal dust emissions 
that originate from the Milky Way. An important method for removing foregrounds  
from observed CMB maps is the so-called internal-linear-combination (ILC) 
technique~\citep{Bennett1992, Tegmark1996, Bennett2003, Tegmark1996, Gold2011}. 
The method was  applied on WMAP observations to estimate the CMB 
temperature anisotropy angular power spectrum by~\cite{Saha2006} and subsequently 
extensively studied~\citep{Saha2008} and applied on WMAP and Planck observations 
~\citep{Saha2017} following a new  improved version, the iterative ILC algorithm in 
harmonic space which  removes  a foreground leakage signal in the final cleaned 
map.~\cite{Saha2018} proposed and implemented a new ILC algorithm on large angular scales
of the sky on CMB temperature anisotropy observations of WMAP and Planck using prior information 
of CMB covariance matrix. This method, by effectively suppressing dominant CMB-foregrounds chance correlations  
at large angular scales, reconstructed a cleaned map and its angular power spectrum accurately 
without any bias.~\cite{Saha2018a} developed a new model independent method, 
the Gibbs ILC method, to 
estimate the joint CMB posterior density and CMB theoretical angular power spectrum 
given the observed foreground contaminated CMB anisotropy data at large angular scales of the sky. 
They provided the best fit estimates of both, CMB temperature map and theoretical angular power 
spectrum along with their confidence interval regions which can directly be 
integrated to cosmological parameter estimation process. In another article~\cite{Saha2020} 
the authors have investigated in detail the impact of random residual error in calibration coefficients 
corresponding to observed CMB maps on the Gibbs ILC estimates at large angular scales.  
 
The ILC method is very appealing  since the weights that are used for linear combinations 
of the input foreground contaminated maps are obtained by a simple analytical expression. 
The advantage of the ILC method is that in order to reconstruct the CMB map one does not 
require to model foreground components, neither in terms of their frequency spectrum nor in 
terms of any templates that trace the  morphological pattern of a foreground components across the sky. This 
greatly reduces the complexity which is otherwise involved in a complete component 
reconstruction  technique, in which 
all foregrounds and CMB component are required to be simultaneously and explicitly modelled in order to 
even reconstruct just the single CMB component. Since the foreground modelling is not 
necessary in an ILC method the cleaned CMB map and its angular power spectrum are, therefore,  not affected 
by any error induced by the uncertainties in the assumed foreground model. Although a significant 
amount of research have already been performed to reconstruct CMB temperature anisotropy signal using 
the  foreground model independent  method its performance on CMB polarization signal reconstruction 
remains yet to be evaluated. During the CMB temperature analysis at large angular scales~\cite{Saha2018} 
reported a large error while reconstructing the CMB signal using usual ILC method. A detailed 
investigation revealed that it is due to the presence of CMB-foreground chance correlation at large 
angular scales and the usual ILC weights were not able to minimize this chance correlation power 
during linear combination. The authors then proposed a new method, the global ILC method~\citep{Saha2018}~\footnote{
The  name `global' is used to indicate that the prior global information from the CMB pixel-pixel covariance 
matrix is used in this algorithm.}, as a remedy 
since incorporating the theoretical CMB covariance matrix in the ILC algorithm effectively 
suppressed these chance correlations and aided in obtaining an accurate CMB map at large angular 
scales. A similar investigation needs to be carried out in the context of foreground removal in 
the CMB polarization analysis. Since CMB polarization signal is much weaker as compared to the CMB 
temperature, it is yet unclear what is the the role of CMB-foreground chance correlation while 
reconstructing a CMB polarization signal at large angular scales and one should, therefore, study 
in detail. In this article,  we seek to obtain an answer to the question - what is the effect of 
CMB-foregrounds chance  correlations at large angular scales on the cleaned CMB E mode map estimated 
using usual ILC method  and how well the  global ILC method  
reconstructs weak CMB E mode signal over large angular scales of the sky using prior information of 
CMB covariance matrix.

A study regarding the performance of the ILC method on CMB E mode reconstruction is necessary also due 
to a different reason. 
A new horizon in cosmology is expected to be seen  with the observations from future generation satellite 
missions which can measure CMB polarization signal with sufficiently large signal to noise 
ratio to measure the weak CMB polarization signal over a wide range of angular scales. Future 
generation CMB polarization observations {\it LiteBIRD}~\citep{Matsumura2014, Sugai2020} 
, Cosmic Origin Explorer (COrE)~\citep{Core2018}, Probe of Inflation and Cosmic Origin 
(PICO)~\citep{PICO2019} have already been proposed. With these sensitive CMB polarization
missions in the making it is again an important question to ask how accurate does the  CMB 
reconstruction techniques perform  
on the simulated observations of these experiments . In this article, we apply the usual ILC and 
global ILC method on the E mode signals of simulated polarization observations of COrE and 
study the performance of each methods.  

For CMB component reconstructions several methods have been proposed in the literature.~\cite{Eriksen2006} proposed a parametric method for CMB and other foreground component 
separation. CMB component and its angular power spectrum along with  foreground components
are reconstructed by~\cite{Eriksen2004, Eriksen2008, Eriksen2008a, Planck_diffuse2018, 
Planck_CMB2016, Bmode2018, CoreB2018} by employing a Gibbs 
sampling technique.~\cite{Saha2011} estimates a foreground cleaned CMB temperature anisotropy map 
using WMAP observations by minimizing a measure of non Gaussianity of the foreground 
contaminated maps. A generalized ILC algorithm for foreground component separation is proposed 
by~\cite{GNILC2011}. \cite{Basak2012} and~\cite{Basak2013} use an ILC method in the needlet 
space.~\cite{Saha2016} use a perturbative technique in the ILC algorithm first proposed by 
~\cite{Bouchet1999} to jointly estimate CMB and foreground 
components in presence of varying spectral index of synchrotron component using simulated 
polarization observations of WMAP  and Planck over large angular scales of the sky.

We organize our paper as follows. In Sec.~\ref{formalism} we discuss the basic formalism 
of this work. We discuss the input frequency maps used in the Monte Carlo simulations 
in Sec.~\ref{input}. We present the results of these simulations in Sec.~\ref{result}. 
We discuss the role of prior in  the global ILC algorithm of this work in Sec.~\ref{role}. 
Finally, we discuss and conclude our results in Sec.~\ref{Con}.  

\section{Formalism} 
\label{formalism}

Since Thomson scattering can produce only linear polarization CMB polarization anisotropy is  
expressed in terms of Stokes  $Q$ and $U$ parameters, which also are the observables 
for a CMB polarization experiment. Since the CMB  polarization fields,  $Q(\hat n)$ and $U(\hat n)$
are defined on the surface of a sphere  and their suitable combinations of full-sky observations  
behave as spin $\pm 2$ functions under a transformation of a local coordinate system, 
these combinations can  be expanded in terms of the corresponding  spin two-basis functions following,
\begin{eqnarray}
Q(\hat n) \pm  iU(\hat n) = \sum_{\ell=2}^{\ell_{max}}a_{\pm 2, \ell m} Y_{\pm 2, \ell m}(\hat n) \, , 
\end{eqnarray} 
where $Y_{\pm 2, \ell m}(\hat n) $ represents the spin $\pm 2$ spherical harmonics and $\hat n$ 
represents a direction vector. Using the spin 2 spherical harmonic coefficients, $a_{\pm 2, \ell m}$
one can define
the spin-0 polarization map, $E(\hat n)$ following
\begin{eqnarray}
E(\hat n) = \sum_{\ell =2}^{\ell_{max}}a^E_{\ell,m} Y_{\ell m}(\hat n)\, , 
\label{qu2e}
\end{eqnarray} 
where $a^E_{\ell m} = \big( a_{2,\ell m} + a_{-2,\ell m}\big)/2$~\footnote{In a similar fashion 
the B-mode map can be defined as $B(\hat n) = \sum_{\ell =2}^{\ell_{max}}a^B_{\ell m} Y_{\ell m}(\hat n)$,
where  $a^B_{\ell m} = \big( a_{2,\ell m} - a_{-2,\ell m}\big)/2i$.}.

Let us assume that  we have full-sky observations of CMB polarization at $n$  different frequency maps. 
The net E mode signal at frequency $\nu_i$ in thermodynamical temperature unit is given by, 
\begin{eqnarray}
S^i(\hat n) = S_0(\hat n) + F^i(\hat n) \, , 
\label{NET_E} 
\end{eqnarray}
where $S_0(\hat n)$ represents the CMB signal, which is independent on frequency $\nu$ due to black-body
nature of CMB~\citep{Mather1994} and $F^i(\hat n)$ denotes the total foreground emission at the frequency $\nu_i$~\footnote{In real experiments 
and for finite pixelation of the sky both the CMB  and foreground signals are smoothed by the beam and pixel 
window functions in Eqn.~\ref{NET_E}, which  in general depend on the frequency $\nu_i$.  In Eqn.~\ref{NET_E} 
we assumed that all frequency maps are already smoothed by a common beam and pixel window functions and hence, omitted
any explicit reference of these functions. For a discussion of common beam and pixel window functions compatible with
this work we refer to Section~\ref{input}.}. We have not included 
any detector noise contribution in Eqn.~\ref{NET_E} since for observations like COrE the detector noise 
level, after $4$ years of its observations, is negligible compared to expected level of CMB and foreground 
signal. We, however, note that the CMB E mode signal reconstruction method described in this work, does incorporate the 
small level of detector noise compatible to COrE. Using all $n$ available frequency maps,  cleaned E mode CMB map can be formed following the usual ILC approach in
the pixel space, 
\begin{eqnarray}
X(\hat n) = \sum_{i=1}^nw_iS^i(\hat n) \, , 
\label{cmap}
\end{eqnarray}
where following the usual ILC method, the weight factors $w_i$ can be  found by minimizing the variance ($\sigma^2)$ of the cleaned map defined 
as 
\begin{eqnarray}
\sigma^2 = {\bf X}^T{\bf X} \,,
\label{var}
\end{eqnarray} 
where, ${\bf X}$ represents $N \times 1$ column vector representing the cleaned map. $N$ denotes the total
number of pixels in the map. Using Eqns.~\ref{cmap}
in Eqn.~\ref{var} we obtain, 
\begin{eqnarray}
\sigma^2 = {\bf W}{\bf A}{\bf W}^T\, , 
\label{red_var1}
\end{eqnarray}
where ${\bf W}$ is $1\times n$ row vector containing the weights and $(i, j)$ element of $n \times n$ 
matrix ${\bf A}$ is given by,
\begin{eqnarray}
{A}_{ij} = {\bf S}^T_i{\bf S}_j\, . 
\label{Aij} 
\end{eqnarray}
Since CMB follows black body spectrum with a very good accuracy  its polarization anisotropy are independent 
on frequency implying elements of its projection vector (also known as shape vector ${\bf e}$) in all frequency bands are unity
in any chosen scale (i.e., {\bf e} = \{1, 1, ...\} -  a row vector of size  $n$). In order to reconstruct the CMB 
component without any normalization bias we must therefore have $\sum_{i=1}^nw_ie_i ={\bf We}^T= 1$.  Minimizing 
$\sigma^2$ from Eqn.~\ref{red_var1} subject to the constraint  on the weights one  obtains,
\begin{eqnarray}
{\bf W} = \frac{{\bf e A}^{\dagger}}{{\bf e A}^{\dagger}{\bf e}^T} \, . 
\label{weight} 
\end{eqnarray}
where $\dagger$ represents the Moore Penrose generalized inverse matrix~\citep{Penrose}.

If detector noise level is negligible and one has sufficient number of input frequency maps so 
that all foreground components can be removed the cleaned  E mode map will have a covariance 
structure consistent with the expected theoretical covariance matrix ${\bf C}$. An improved global ILC 
method can therefore be constructed by demanding this condition on the cleaned E mode map~\citep{Saha2018}. This
can be achieved by minimizing CMB covariance weighted 
 variance ($\sigma^2_{\tt red}$) of the cleaned map instead of usual variance. In the context of 
 CMB E mode analysis,  the CMB weighted variance is defined as follows 
\begin{eqnarray}
\sigma^2_{\tt red} = {\bf X}^T{{\bf C}}^{\dagger}{\bf X} \,,
\label{red_var}
\end{eqnarray} 
where, ${\bf C}^{\dagger}$ represents $N \times N$ Moore-Penrose generalized 
inverse of CMB E mode covariance matrix ${\bf C}$ in pixel space~\footnote{We note that ${\bf C}$ is a singular matrix since 
it has a total number of independent degrees of freedom $(\ell_{max}+1)(\ell_{max}+2) (\ell_{max}+1)-4$ which is less 
than its size $N = 12 \times N^2_{side}$ where $\ell_{max} = 2N_{side}$. Therefore, in Eqn.~\ref{red_var} we use
Moore-Penrose generalized inverse instead of the usual inverse. }. Assuming CMB E mode
signal is statistically isotropic, $(i,j)$  element of ${\bf C}$  follows, 
\begin{eqnarray}
C_{ij}  = \sum_{\ell =2}^{\ell_{max}}\frac{2\ell +1}{4 \pi} C^E_{\ell}\mathcal P_{\ell}(\hat n_i\cdot \hat n_j)B^2_{\ell}P^2_{\ell}\, , 
\end{eqnarray}   
where $C^E_{\ell}$ denotes the theoretical CMB angular power spectrum and  $\hat n_i$ $(\hat n_j)$ is the direction corresponding 
the pixel $i$ $(j)$. $B_{\ell}$ and $P_{\ell}$ are respectively the common beam and pixel window functions for our analysis. 
Using Eqns.~\ref{cmap} and~\ref{red_var}, the $(i,j)^{\tt th}$ element of the new {\bf A} matrix is 
obtained as
\begin{eqnarray}
{ A}_{ij} = {\bf S}^T_i{\bf C}^{\dagger}{\bf S}_j\, ,  
\label{Aij_new} 
\end{eqnarray}
which is then substituted in Eqn.~\ref{weight} to obtain the weights.

In the pixel space computing elements of matrix ${\bf A}$ is computationally intensive 
since it  involves finding the pseudo inverse of the pixel space E mode covariance matrix which requires 
$N^3$ operations. However, as shown in~\cite{Saha2018} 
the elements $A_{ij}$ of the matrix ${\bf A}$ (Eqn.~\ref{Aij_new}) can be computed efficiently in the multipole space using  
\begin{eqnarray}
A_{ij} = \sum_{\ell =2}^{\ell_{max}}\left(2\ell+1\right) \frac{\sigma^{ij}_{\ell}}{C^E_{\ell}B^2_{\ell}P^2_{\ell}}\,, 
\label{Aij1} 
\end{eqnarray}   
where $\sigma^{ij}_{\ell}$ represents the cross-power spectrum of the frequency map ${\bf S}_i$ and ${\bf S}_j$. 
These cross spectra are already smoothed by the beam and pixel window functions.  

\section{Input Frequency maps}

\label{input}

\begin{table}
	\caption{ List of $15$ COrE frequencies used for CMB E mode reconstruction shown in first column. The second column 
   represents the FWHM of circularly symmetric Gaussian beam functions of COrE detectors. The last column shows the noise 
levels $\sigma_i$ for the frequency $\nu_i$.}
	\hspace{7em}	\begin{tabular}{lll}
		\hline
		Frequency &  Beam FWHM & $\Delta Q=\Delta U$            \\
		(GHz)     &   (arcmin)     & 
		($\mu$K.arcmin)           \\   
		\hline
		60   & 17.87 & 7.49 \\
		70   & 15.39 & 7.07  \\
		80   & 13.52 & 6.78 \\
		90  &  12.08 & 5.16 \\
		100 &  10.92 & 5.02 \\
		115  & 9.56 &  4.95  \\   
		130 &  8.51 &  3.89 \\
		145 &  7.68  &  3.61 \\
		160 &  7.01  &  3.68\\ 
		175 &  6.45  & 3.61\\
		195 &  5.84  &  3.46 \\
		220 &  5.23  &  3.81 \\
		255 &  4.57  &  5.58 \\
		295 &  3.99  &  7.42 \\
		340 &  3.49   &11.10  \\
		\hline
	\end{tabular}
         \label{table}
\end{table}

We simulate realistic Stokes Q and U polarization sky maps containing CMB, foreground and detector noise 
 as will be observed by COrE at  $15$ frequencies between 
$60$ GHz and $340$ GHz. We do not use higher frequency bands since they are intrinsically more noisy. 
Frequencies corresponding to different COrE frequency maps used in this work are shown in Table~\ref{table}. 
Since we are interested in reconstruction of CMB E mode over large angular scales we convert the Stokes Q, U maps 
to E mode polarization maps (following Eqn.~\ref{qu2e}) at HEALPix~\footnote{Hierarchical Equal-Area IsoLatitude Pixallation of sphere (e.g.,~\cite{Gorski2005}).} pixel resolution $N_{side} = 16$ and beam resolution of a Gaussian 
beam of full-width at half-maximum (FWHM) of $9^\circ$.  The choice of this beam smoothing corresponds to 
roughly $2.5$ times the pixel width at $N_{side} = 16$ and results in band-width limited signal in all input frequency maps. 

We simulate a total of $1000$ (random) realizations of input frequency E mode CMB maps corresponding to the 
frequency maps of our interest.  To simulate the CMB E mode signal we first generate the CMB stokes Q, U 
maps at $N_{side} = 16$ and FWHM = $9^\circ$ with necessary pixel-window smoothing taken into account
using CMB E mode best-fit theoretical angular power spectrum~\citep{PlanckCl2018}. 
We convert the Stokes Q, U maps to E mode maps at the same beam and pixel resolution following Eqn.~\ref{qu2e}.  
For polarization anisotropy two major foreground components are synchrotron and thermal dust. We simulate 
these components at our beam and pixel resolution following a method similar to~\cite{Bmode2018} and again 
convert them to E mode signals at $15$ COrE frequencies of Table~\ref{table}. For synchrotron and thermal dust we 
use a rigid frequency spectrum model of $\beta_s =  -3.0$ and $\beta_d = 1.6$ respectively. We simulate the 
detector noise Stokes Q and U maps with polarization noise levels compatible to $4$ years of COrE observations (e.g., Table 1 
of~\cite{Core2018}). Denoting the (Q or U) noise standard deviation of a frequency $\nu_i$ of~\cite{Core2018}
as $\sigma_i$ (in thermodynamic $\mu K.{\textrm {arcmin}}$ unit), the noise standard deviation at $N_{side} = 16$ for Stokes Q
or $U$ maps, without any beam or pixel smoothing effects taken into account, is given by, $\sigma^0_i = \sigma_ic/\Delta_{pix}$
where $c$ denotes the conversion factor from arcmin to radian unit, and $\Delta_{pix}$ denotes width of an $N_{side} = 16$ pixel
in radian. The noise standard deviations, $\sigma_i$, are mentioned in the third column of Table~\ref{table}. We simulate random realizations of 
Stokes Q and U noise maps which are uncorrelated with frequencies (and also between themselves) by multiplying these
standard deviation values with independent normal deviates corresponding to Gaussian beam resolution of FWHM = $9^\circ$.
~\footnote{In a given (high resolution) input frequency map obtained by some map making algorithm from time-ordered 
data the detector noise component is not smoothed by the beam or pixel window functions. However, for a low 
resolution analysis a strategy for downgrading the maps from the higher resolution is to first convert the
frequency maps in harmonic  space, getting rid of high resolution beam and pixel window functions, smoothing by the 
new beam and pixel window function corresponding to chosen  low resolution and finally convert the resulting smoothed 
spherical harmonic coefficients to map domain. This process inevitably smooths  the noise of a particular frequency $\nu_i$ 
of the low resolution maps by the effective window function $W^i_\ell= B^0_\ell P^0_\ell/B^i_\ell P^i_\ell$, where $B^0_\ell$
and $P^0_\ell$ are respectively beam and pixel  window functions  of the low resolutions maps and the corresponding quantities 
with superscript $i$ are respectively beam and pixel window functions of the high resolution maps. In this work, the 
noise maps at the low resolution are merely smoothed by the ratio $W^i_\ell= B^0_\ell/B^i_\ell$, without taking into account any pixel
window effects. We assume $B^i_{\ell}$ are beam window functions for circularly symmetric Gaussian beam functions which 
are tabulated in the second column of Table~\ref{table}.  The pixel window smoothing effect is expected to be negligible at $N_{side} = 16$ given 
the high beam resolutions of COrE frequency maps, which requires a finer pixel grid for the high resolution analysis. }
Finally, we convert smoothed Q, U noise maps  to
E mode noise maps following Eqn.~\ref{qu2e}. We add synchrotron and thermal foreground E mode maps 
and randomly generated E mode noise maps corresponding to different frequency  maps with a given CMB E mode 
map  to  simulate 
a single set of COrE E mode frequency map. We generate a total of $1000$ random realization of E mode frequency 
maps for Monte Carlo simulations of CMB E mode reconstruction. Detector noise signals are random and uncorrelated 
between any two of these input sets. Foreground components are fixed without any randomness in all sets of 
simulations.        

An important point to note that while converting the Stokes Q and U maps to E mode maps for any component
we use the conversion over the fullsky. This avoids any problem of leakage of E mode signal to B mode
signal and vice versa which results in case of any such conversions over the partial sky~\citep{Lewis2002, Lewis2003}.   
 
\section{Results}
\label{result}
\begin{figure}
\vspace{-4cm}
  \hspace{3cm}
        \includegraphics[scale=1.3]{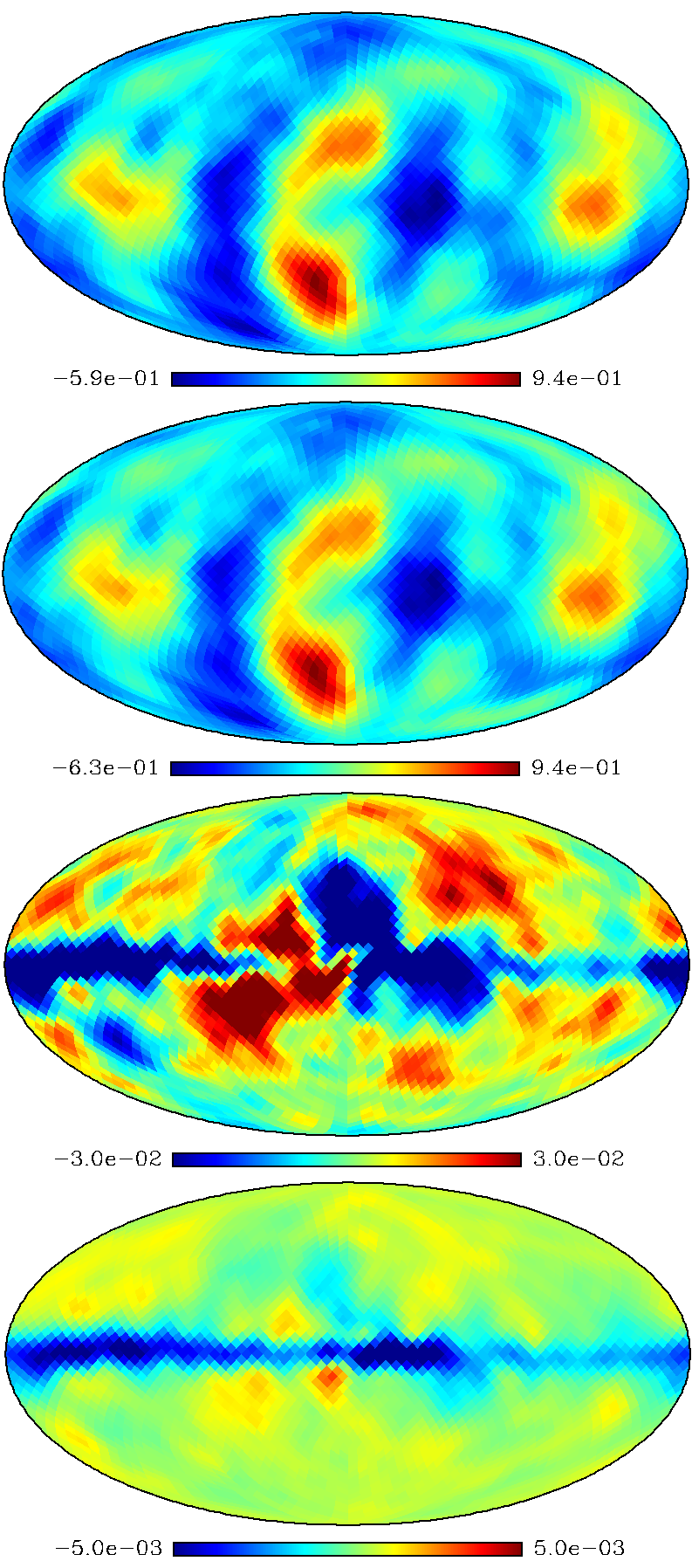}
        \caption{Top panel shows  one random realization (seed = 100) of  CMB E mode map at $N_{side} = 16$ and 
           Gaussian beam resolution of FWHM = $9^\circ$. The second panel shows the cleaned E mode map obtained 
          following usual ILC algorithm (without using 
                any prior information about the CMB covariance matrix) for the randomly chosen pure CMB E map 
                shown in top panel. The third panel shows the difference between the second panel
                and the input CMB E mode map. A strong level of residuals is seen in the difference map. The bottom 
                panel shows the mean of $1000$ difference maps formed by subtracting the pure CMB map from cleaned map 
                following usual ILC method. The figure shows presence of residual contamination within the 
                pixel range shown.} 
        \label{cmap_ilc}
\end{figure}

In this section, we discuss the results from the detailed Monte Carlo simulations in which  we  simulate
$1000$ different set of input foreground contaminated CMB E mode maps at all COrE frequencies. We implement
 usual ILC and global ILC method to reconstruct the corresponding E mode CMB signal after properly minimizing 
the foregrounds in  foreground model independent manner.
The main scope of this section is to investigate in  CMB E mode reconstruction over the large angular scales 
of the sky how significant are the the bias and errors in the cleaned CMB E mode map, corresponding CMB angular power 
spectrum obtained following usual ILC method and  the  global ILC method  respectively.
\vspace{3cm}
\subsection{Usual ILC Method}
\label{UsualILC}
\begin{figure}
\hspace{2cm}
	\includegraphics[scale=1.5]{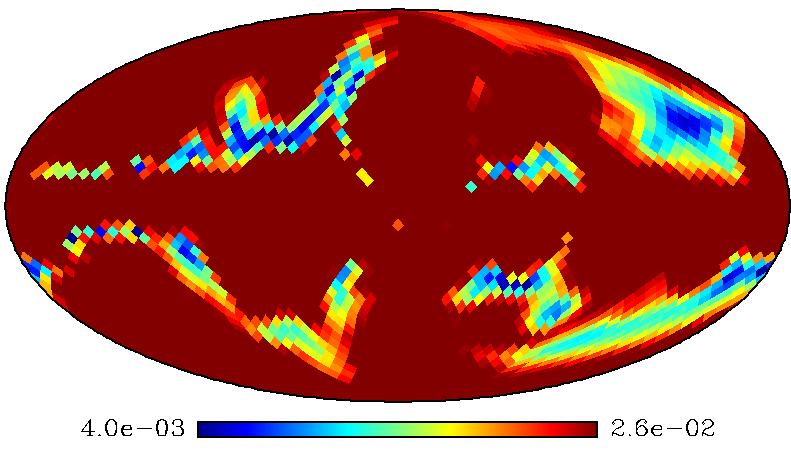}
	\caption{ Figure showing standard deviation map computed from $1000$ difference 
		maps obtained by subtracting the input CMB E maps from the corresponding 
                cleaned E maps for usual ILC method. Poor reconstruction accuracy in usual ILC over large angular scales 
		is readily concluded from this figure.  }
	\label{cmap_ilc1}
\end{figure}

In the top panel of Fig.~\ref{cmap_ilc} we show the input E mode CMB map for a randomly chosen 
seed $100$. Corresponding cleaned E mode map obtained by using the usual ILC method is shown 
in the second panel of this figure.  The color-scale of these subfigures are 
 in $\mu K$ (thermodynamic) temperature.  Although, both these subfigures look similar 
there are differences between the actual pixel distributions in each case, which may be inferred from the values 
indicated by the corresponding color scales. In the third  panel of Fig.~\ref{cmap_ilc}  we show the 
difference between the cleaned  E map  obtained by the usual ILC and the corresponding input E mode 
CMB map for the random seed $100$.  Residuals of complex structure is prominently visible in the difference map in the plotted 
color scale showing significant residuals may exist of magnitude even as large as $0.03 \mu K$ in the case 
of usual ILC method where no prior information about the CMB E mode covariance matrix is used. Using 
all $1000$ E mode cleaned maps obtained by the usual ILC method we estimate the mean difference map of
cleaned and input CMB E mode signal. The resulting map is shown in the fourth panel of Fig.~\ref{cmap_ilc}. 
  The standard deviation map 
of residuals obtained from the simulations of usual ILC method shows very large errors which is shown in the 
in Fig.~\ref{cmap_ilc1}.

\begin{figure}
\vspace{-2cm}
\hspace{1cm}
	\includegraphics[scale=0.7]{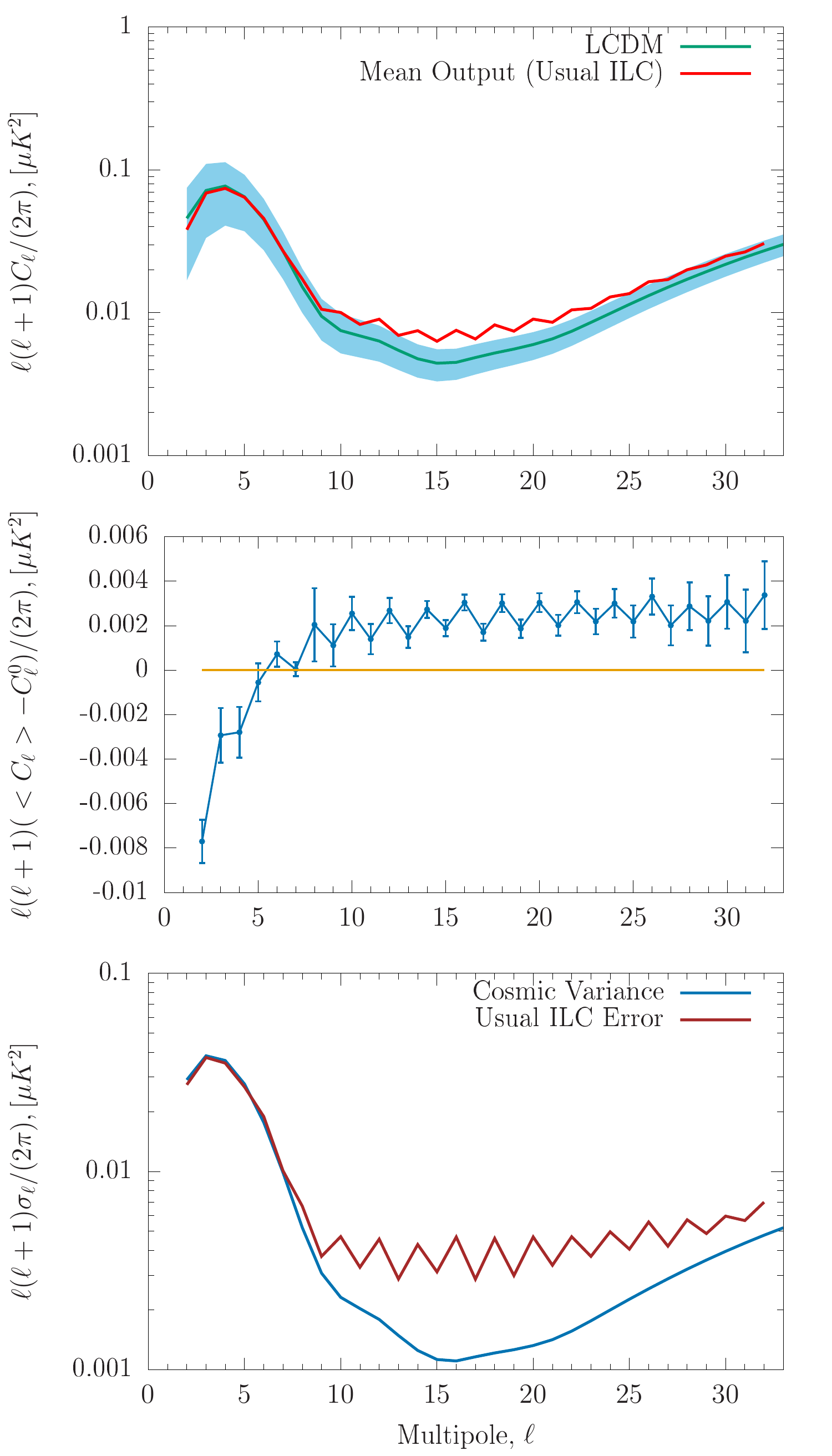}
	\caption{Top panel shows the  mean cleaned E mode angular power spectrum obtained
		by following usual ILC method from $1000$ Monte Carlo simulations in red. The blue 
		line shows the theoretical E mode angular power spectrum along with the theoretical 
		cosmic variance induced error with light blue band. A  negative bias at low 
		multipoles and very strong positive bias (due to foreground residuals) are seen for multipoles 
		starting $\ell \sim 7$. The middle panel shows the bias in the cleaned E mode 
		power spectrum. Error-bars shown are applicable for the mean spectrum and are scaled versions 
		(by $1/\sqrt{1000}$) of errors computed from the global ILC 
		method. For $\ell \ge 8$ error bars are scaled additionally by a factor of $10$ to make 
		them visible on the scales of the plot.  The bottom panel shows a comparison of errors on cleaned  E mode 
		angular power spectrum (brown) and the theoretically computed cosmic variance induced 
		error (blue). Inaccurate CMB E mode reconstruction following the usual ILC method 
		is easily concluded from this plot.}
	\label{cl3}
\end{figure}

Using $1000$  cleaned E mode angular spectra (after removing beam and pixel smoothing effects) 
from the cleaned E mode maps obtained by the usual ILC method we  estimate a mean cleaned E mode
angular power spectrum. The mean spectrum (in red) along with the input theoretical E mode angular power
spectrum (in blue) is shown in the top panel of Fig.~\ref{cl3}. The theoretical cosmic variance 
induced error is shown in this figure in light blue band. At the low multipoles $\ell \lesssim 4$
we see a  negative bias in the cleaned power spectrum which has been also reported earlier
for the CMB temperature anisotropy reconstruction following  usual ILC method~\citep{Saha2006,Saha2008, Saha2017}.
The negative bias decreases with 
increase in $\ell$. Starting from $\ell \sim 7$ a strong residual foreground bias comes into existence 
in cleaned E mode spectrum obtained by this method. The bias in this multipole range becomes at least equal 
to and in several multipoles even more compared to the cosmic variance induced error. To emphasize 
the bias in the cleaned angular power spectrum is maximum around $\ell \sim 15$ where the expected CMB 
E mode signal is the weakest in the multipole range considered in this work. It is interesting to note 
here that the even multipoles has larger power due to residual foregrounds as compared to odd multipoles. 
The middle panel of 
this figure shows the bias in the cleaned E  mode angular power spectrum at different multipole
ranges. The error-bars shown in this panel are computed from the simulations of global ILC 
method, however, they are estimated for the mean spectrum obtained from $1000$ simulations, by multiplying the 
global ILC errors at different multipoles by the factor $1/\sqrt{1000}$.  For $\ell \ge 8$ these
scaled  errors are multiplied by an additional factor of $10$ to make them visible on the vertical scale
of this plot. A highly significant negative bias is observed in the usual ILC E mode angular spectrum
at the low multipoles. A positive bias due to foreground residuals becomes very strong starting from $\ell =8$.  The bottom panel 
of Fig.~\ref{cl3} compares  the error-bars on the cleaned E mode angular spectrum with the cosmic variance 
error. Significantly larger errors compared to the cosmic variance prediction is observed for usual 
ILC reconstruction for the multipole starting from $\ell \sim 7$.    It is interesting to note that
the foreground residuals in the usual ILC method cause even multipoles to be more erroneous 
than the odd multipoles starting from $\ell \sim 9$. The bias for even multipoles in the
cleaned EE mode angular power spectrum similarly is larger
than the bias in neighbouring odd multipoles for $\ell \ge 7$ as
seen from the second panel of Fig.~\ref{cl3}.

\subsection{Global ILC Method} 
\begin{figure}
\vspace{-2cm}
\hspace{1.3cm}
\includegraphics[scale=1.7]{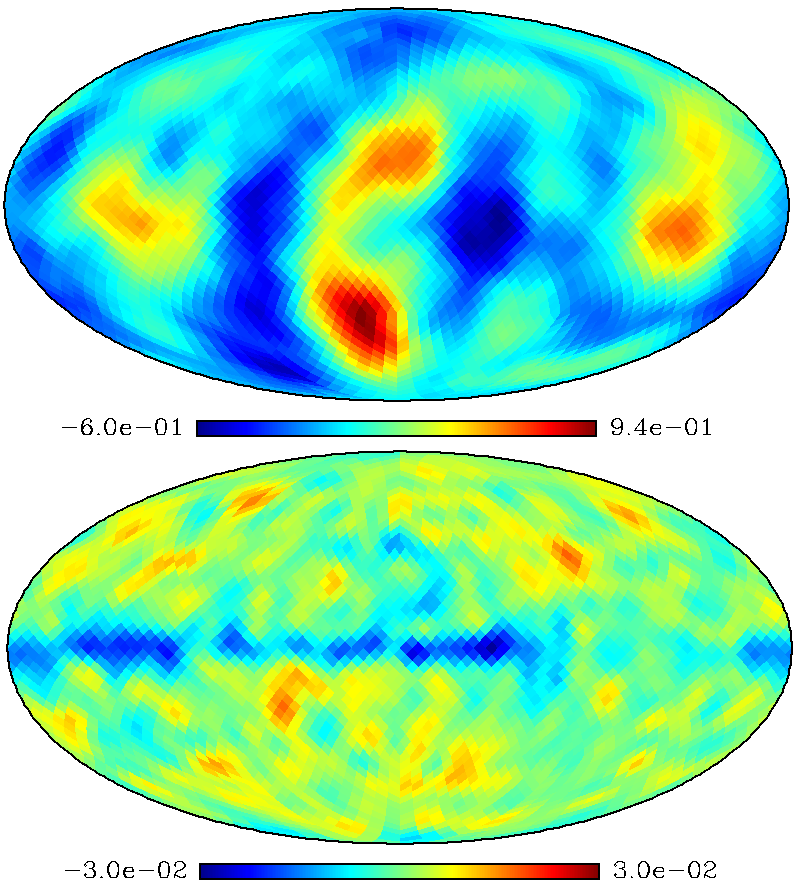}
\caption{Top panel shows the reconstructed CMB E mode map following  global ILC method using the prior 
knowledge of CMB E mode covariance  matrix for the randomly chosen input CMB E mode map shown in top panel
of Fig.~\ref{cmap_ilc}. Both these figures agree well with each other. 
The bottom panel shows the difference between the top panel and the input CMB E mode map (top panel of Fig.~\ref{cmap_ilc})
and hence  represents residuals in global ILC 
method. Units for all maps are in $\mu K$ (thermodynamic).}
\label{cmap1}
\end{figure}
In this section we discuss the results of Monte Carlo simulations of global ILC method, where we  
use prior information about the CMB E mode covariance matrix.
Using each set of input frequency maps we estimate the CMB covariance matrix weighted matrix $A_{ij}$ 
using Eqn.~\ref{Aij_new} and compute the corresponding set of weights following Eqn.~\ref{weight}. We obtain
a total of $1000$ cleaned E Mode CMB maps corresponding to all sets of input frequency maps. Apart from 
the CMB signal each of these cleaned maps contain small levels foreground residuals  and minor level of detector 
noise. We show the input CMB E mode map for a randomly chosen seed ($100$) in top panel of Fig.~\ref{cmap_ilc}. 
As is the case for this figure the
typical pixel values for the large scale CMB E mode signal with the chosen beam resolution of $9^\circ$ 
of this work,  lies within $\sim \pm 1 \mu K$~\footnote{For pure E mode CMB maps at $N_{side}= 16$ and $9^\circ$
Gaussian beam smoothing we find that the  mean standard deviation from $1000$ random CMB realizations is
 $0.28 \mu K$. }. The top panel of Fig.~\ref{cmap1} shows the 
reconstructed  CMB E mode map  using the global ILC method where prior information about the CMB E mode 
covariance matrix is used. Visually, both the input CMB E mode map (top panel of Fig.~\ref{cmap_ilc}) and  
 the cleaned map obtained by the global ILC method agree well with each other. In bottom panel of Fig.~\ref{cmap1} we show the difference
between the cleaned and the input E mode map. A small level of residuals is seen in this figure. 
The foreground residuals of this figure is clearly 
much smaller than the corresponding figure 
for the usual ILC case, shown in third panel of Fig.~\ref{cmap_ilc}

\begin{figure}
\includegraphics[scale=1.9]{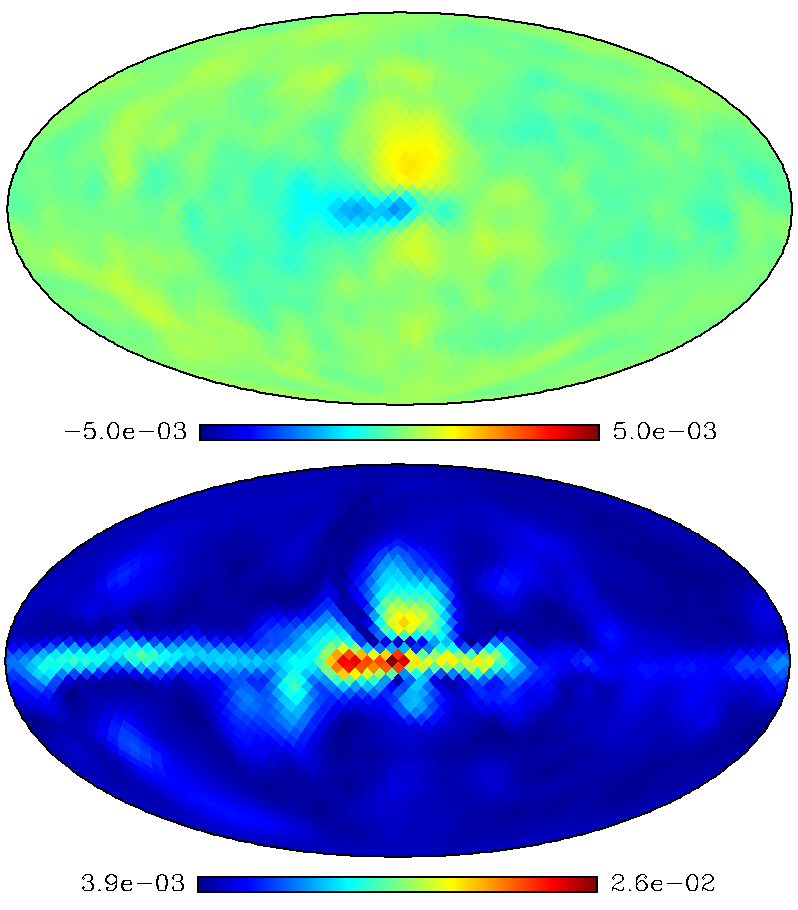}
\caption{Top panel shows mean of $1000$ difference maps computed by subtracting input pure CMB E mode 
polarization maps from the cleaned E mode maps obtained by using global ILC method. The mean map shows 
average foreground residuals (which are $\lesssim 0.005 \mu K$) present in the E mode CMB reconstruction method.
The bottom panel shows the standard deviation map obtained from all $1000$ difference maps. Apart from the 
central and somewhat left side  of the galactic plane the $1\sigma$ reconstruction error sky pixels are $\lesssim 0.004 \mu K$. }
\label{cmap_fig}
\end{figure}

To statistically assess the performance of the global ILC method to reconstruct  the large scale cleaned E 
mode map  we estimate the average foreground residuals in each of the cleaned maps by computing the mean
difference map obtained by subtracting all $1000$ cleaned output and the corresponding input pure CMB E mode maps. The mean difference map is shown 
in the top panel of Fig.~\ref{cmap_fig}. The maximum of absolute pixel values of this map is well less than $0.005 \mu K$
much less than  typical value of pure CMB E mode standard deviation ($\sim 0.28 \mu K$), indicating that the mean residual foregrounds 
in the cleaned maps are less than these
estimates.  The bottom panel of Fig.~\ref{cmap_fig} shows the standard deviation map computed from all the 
difference maps. As seen from this map over most of the regions of the sky the reconstruction error
is only $\sim 0.004 \mu K$. Near the galactic plane the error tends to increase towards the left side
and becomes maximum towards the galactic center where the maximum error becomes $0.026 \mu K$.    

\begin{figure}
\vspace{-2cm}
\hspace{1cm}
\includegraphics[scale=0.65]{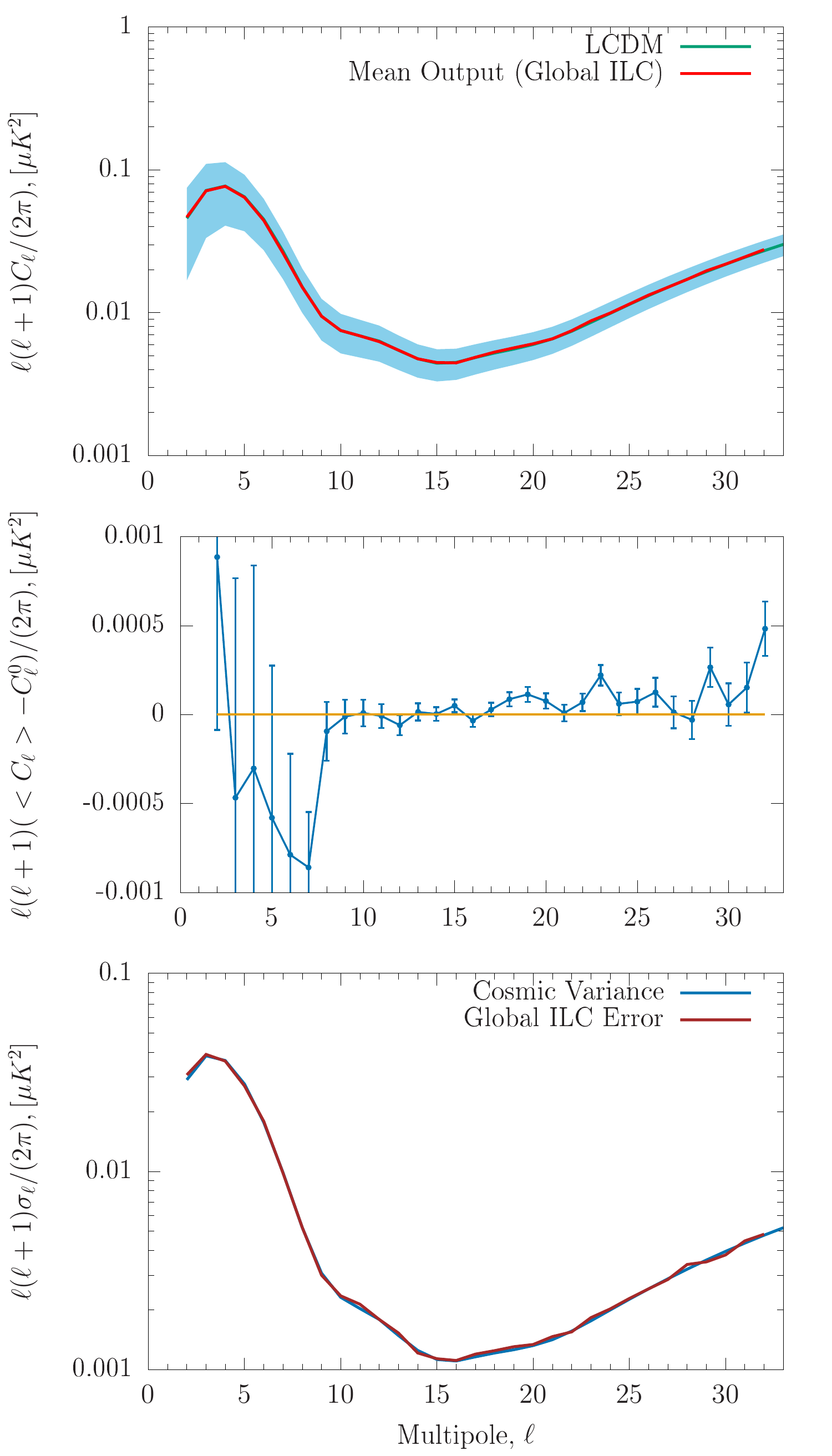}
\caption{Top panel shows mean  cleaned CMB E mode angular power spectrum (in red) obtained from 
global ILC method of this work using all $1000$ Monte Carlo simulations. The theoretical 
E mode angular power spectrum is shown in blue line but lies completely below the mean cleaned 
power spectrum on the scale of this plot. Theoretical cosmic variance  induced error is shown in light blue 
band.  The middle panel shows the difference between the mean cleaned and the theoretical E mode 
angular power spectrum in blue line. The error-bars shown are computed from Monte-Carlo simulations and 
applicable for  the mean cleaned spectrum. The bottom panel shows the comparison of error on cleaned E 
power spectrum and theoretically estimated  cosmic variance induced  error. Both these estimates agree 
very well with each other indicating accurate CMB E mode reconstruction by global ILC method.   } 
\label{cl2}
\end{figure}

We estimate E mode CMB angular power spectrum from each of the $1000$ cleaned maps and remove the beam 
and pixel effects. We plot the average of all these spectra along with the input theoretical EE power 
spectrum used in all the Monte Carlo simulations to generate the CMB E mode map in top panel of Fig.~\ref{cl2}. 
The mean CMB E mode spectrum (red) agrees very well with the theoretical spectrum (blue line), both
of which are indistinguishable on the scale shown in this panel. It is specifically important  to note that the 
global ILC method does not show presence of any positive bias, which may arise due to significant 
foreground residuals in the cleaned maps. Moreover, the average cleaned power spectrum does not show 
any indication of negative bias at the lowest multiples, which has been reported earlier for the case 
of CMB temperature anisotropy reconstruction following usual ILC method~\citep{Saha2006,Saha2008, Saha2017}. This negative bias effect is 
also discussed later in Sec.~\ref{UsualILC} for the case of the usual ILC method in the context of CMB 
E mode reconstruction.  The $1\sigma$ error band due to 
cosmic variance alone is also shown in this panel in blue. The middle panel of this figure shows 
the difference  between the averaged cleaned E mode spectrum and the input theoretical power spectrum 
in blue line. The error-bars plotted in this figure are applicable for the averaged cleaned spectrum and 
is obtained by scaling the standard deviations of the cleaned angular spectrum obtained from the Monte Carlo
simulations by the factor $\sqrt{1/1000}$. Clearly, the mean cleaned angular power spectrum does not deviate 
from the theoretical E mode power spectrum at the entire multipole range $2\le \ell \le 32$ showing  
no signature of bias in the cleaned power spectrum. The bottom panel of Fig.~\ref{cl2} shows comparison 
between the $1\sigma$ error-bars on the cleaned angular power spectrum at different multipoles estimated from Monte Carlo simulations 
(brown line) along with the theoretical estimates of cosmic variance induced error (in blue line). Both
these estimates  match very well with each other. This shows that the global ILC method reconstructs 
cleaned  E mode angular power spectrum accurately as compared to the usual ILC method where the cleaned 
E mode CMB map and corresponding CMB angular power spectrum obtained from the later suffers from bias 
due to CMB-foregrounds chance correlation power dominant at large angular scales.

\section{Role of Theoretical CMB E mode Covariance matrix in  Global ILC Method} 
\label{role}
\begin{figure}
\hspace{2cm}
\includegraphics[scale=1.5]{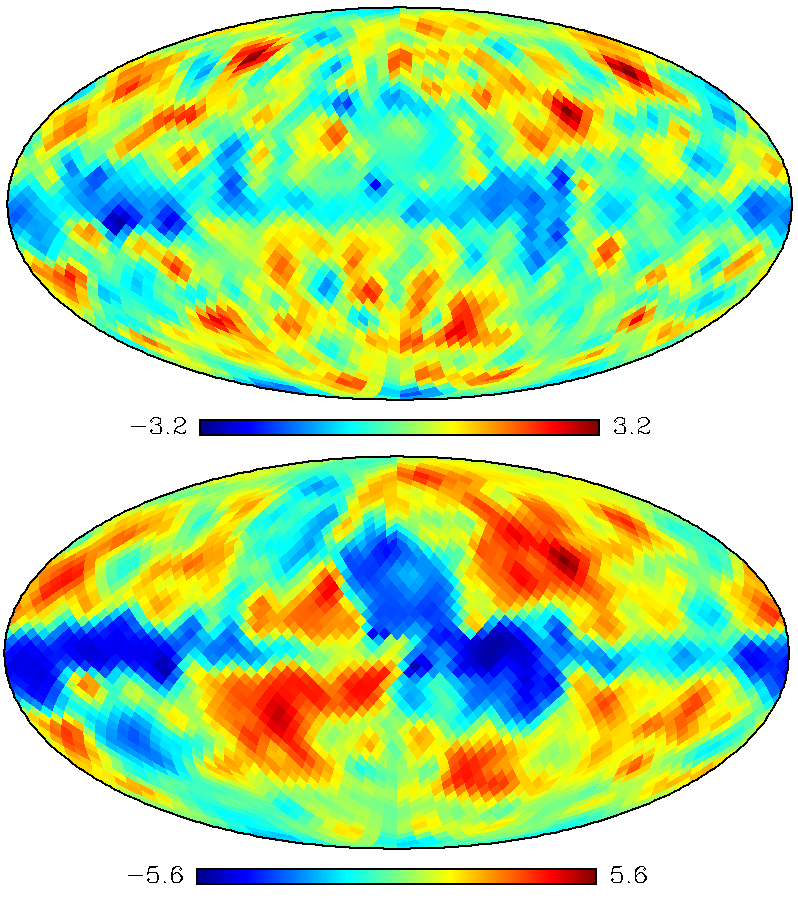}
\caption{Top panel shows the significance of deviation of cleaned E mode 
map obtained following global ILC method from  from the input 
pure CMB  E mode map for a randomly chosen Monte Carlo simulation. The map
represents cleaned output minus input CMB map weighted by the standard deviation 
map all from the global ILC method. The bottom panel shows the same for the 
usual ILC CMB E mode reconstruction, with standard deviation map used for weighting
given also by the global ILC case. The larger significance of deviations 
of the bottom panel is caused by the stronger residuals present in the 
usual ILC case. 
  }
\label{sig_map}
\end{figure} 

\begin{figure*}
\hspace{-2.5cm}
\includegraphics[scale=0.48]{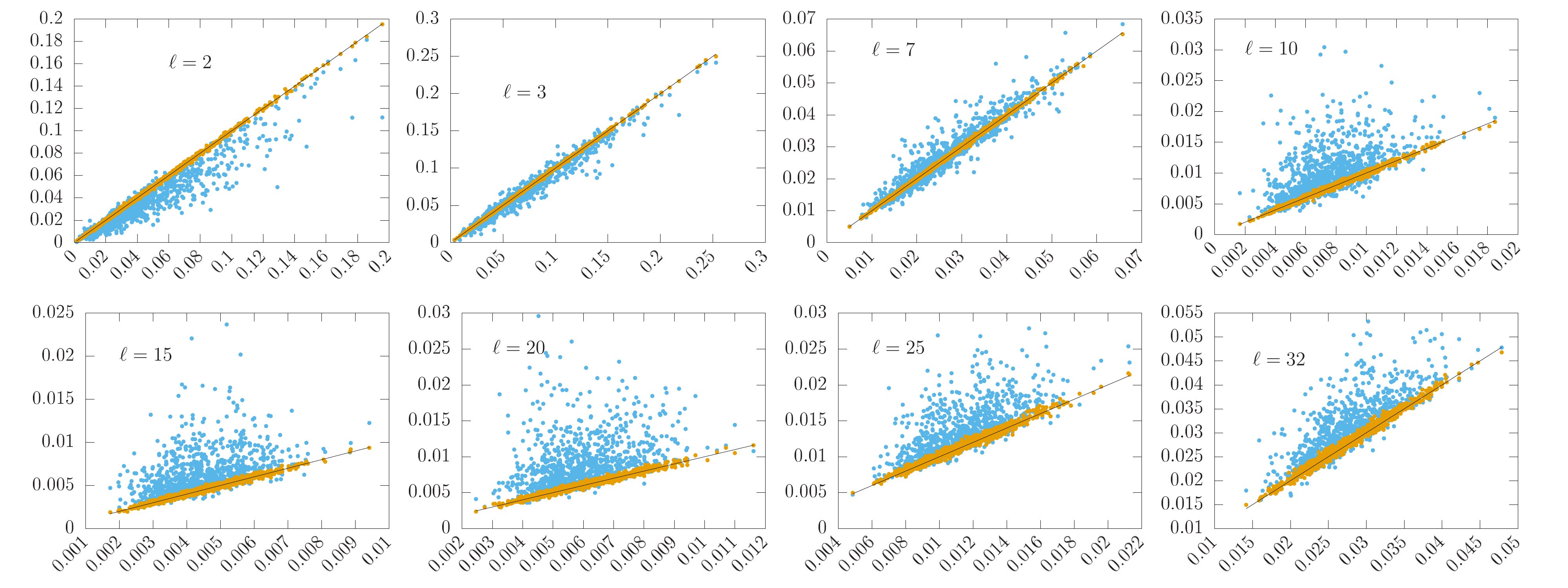}
\caption{Figure showing scatter plot of E mode angular power spectra. The horizontal axis 
of each subplot denotes pure CMB E mode power spectrum for all $1000$ Monte Carlo simulations 
for different multipoles as indicated. Vertical coordinates of golden points are given 
by the cleaned angular power spectra obtained following global ILC method for all $1000$ cases 
for the respective multipoles. The vertical 
coordinates of blue points are given by the cleaned spectra from usual ILC method.  The black 
line represents the straight line with slope unity. Presence of a negative bias at low multipoles and a positive bias
at higher multipoles are evident for the reconstruction following usual ILC method. All axes represent 
$\ell(\ell + 1)C^E_{\ell}/2\pi$ in unit of $\mu K^2$.    }
\label{corr_cl}
\end{figure*}

\begin{figure}
\vspace{-3cm}
\includegraphics[scale=0.9]{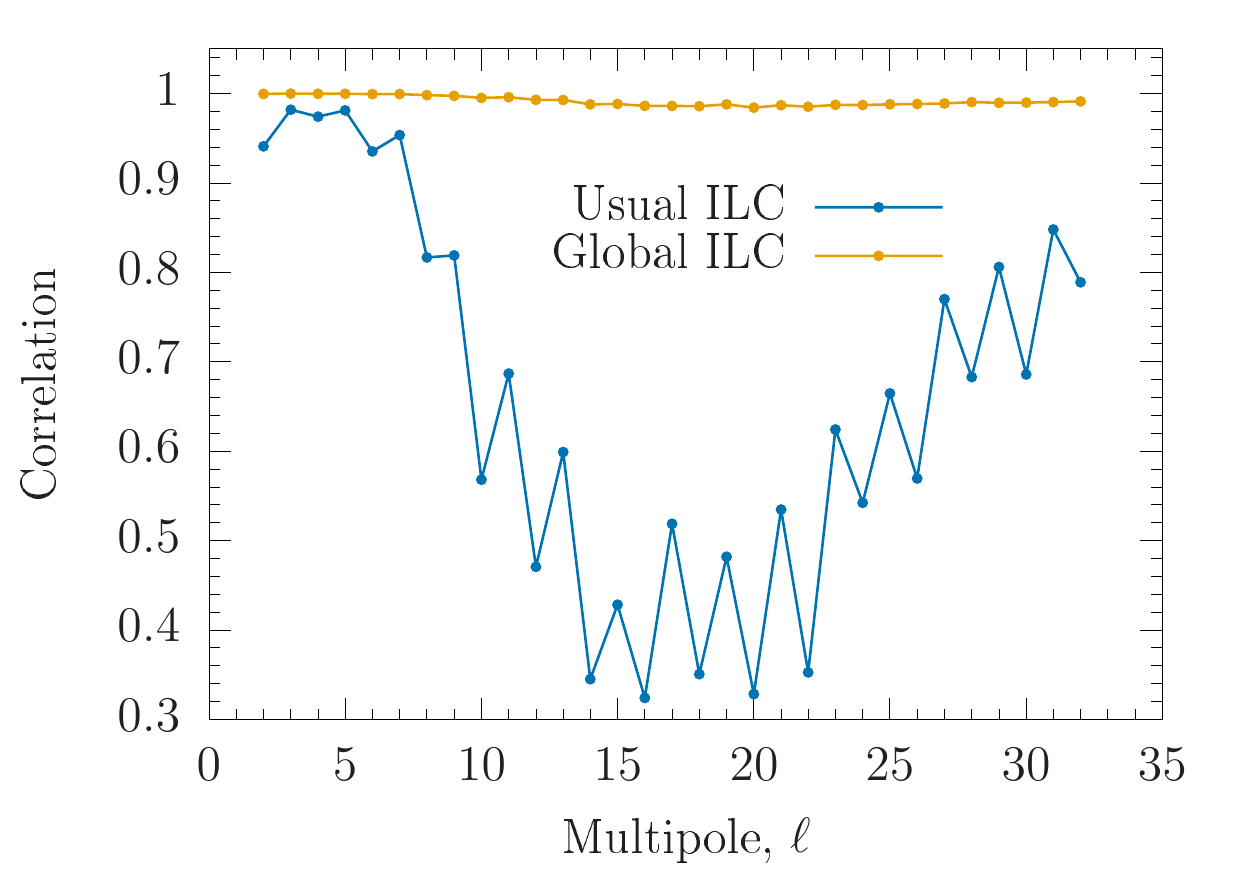}
\caption{Figure showing strong correlation of CMB E mode angular power spectrum obtained by the 
global ILC method (in gold) with the input pure CMB E mode angular power spectrum for the multipole 
range $2 \le \ell \le 32$. The blue line (with points) shows corresponding results for the usual
ILC method. A poor correlation for the usual ILC method and excellent correlation for the global 
ILC method of this work are observed. }
\label{corr_val}
\end{figure}

\begin{figure*}
\hspace{-2cm}
	\includegraphics[scale=0.6]{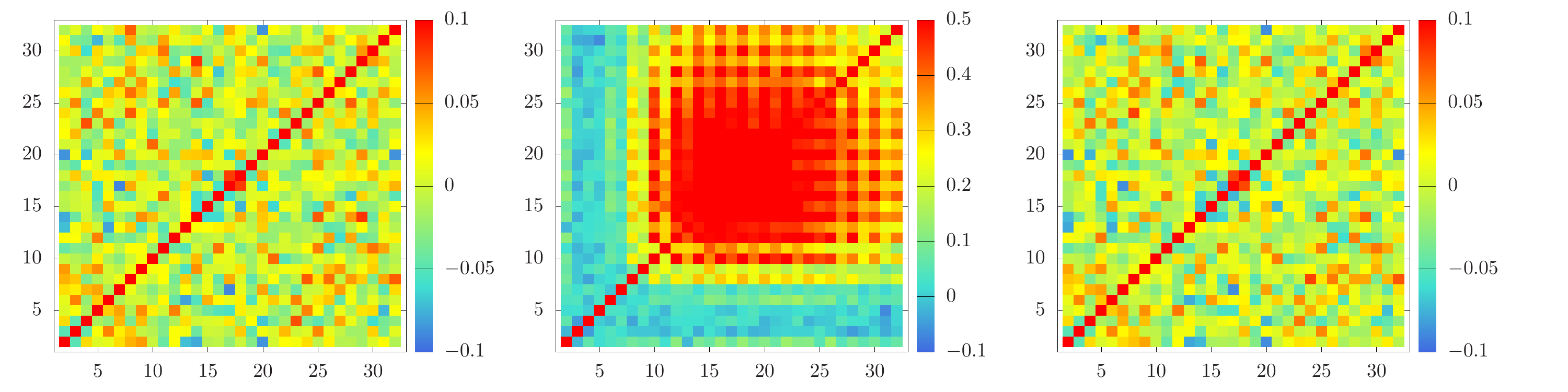}
	\caption{Figure showing the correlation matrix estimated using the angular power 
	spectrum from $1000$ input CMB E mode maps (left panel), cleaned CMB E mode maps after removing 
	foregrounds using usual ILC method (middle panel) and using global ILC method (right panel). 
	We see the correlation matrix estimated from global ILC method is consistent with the input E mode 
	CMB maps. 
         For the usual ILC method a  strong correlation between multipoles due 
	to residual foreground contaminations is visible.}
        \label{corr1}
\end{figure*}	

The chance correlations between CMB and non-CMB E mode signals is a significant source of concern 
for application of ILC method towards  cleaned CMB E mode signal estimation, specifically 
over the large angular scales over the sky. The global ILC method provides a pathway to accurately
reconstruct the CMB signal at the low multipole regions using prior information about the CMB 
covariance matrix, so that the cleaned E mode map has a covariance structure  consistent with the 
expected theory of the CMB angular power spectrum. Using the theoretical CMB covariance matrix model 
the global ILC method effectively suppresses the CMB and non-CMB components chance correlation power at the 
low multipoles through Eqn.~\ref{Aij1} by weighting down the chance-correlation fluctuations
in $\sigma^{ij}_\ell$ by the inverse of E mode theoretical power spectrum. This leads to 
significantly improved estimation of both E mode cleaned map and its angular power spectrum 
by the global ILC method compared to the usual ILC case. 

In top panel of Fig.~\ref{sig_map} we show 
difference of a randomly chosen cleaned E map (corresponding to the seed $100$)  of global ILC 
method and input CMB E map after weighting each pixel 
value by the standard deviation map obtained in the same  method (shown in bottom panel of Fig.~\ref{cmap_fig}).
The pixel values of this weighted difference map represents significance of   deviations, which lie 
within $\pm 3.2 \sigma$, indicating no significant bias present in the cleaned E mode map obtained 
by the global ILC method. In the bottom panel of Fig.~\ref{sig_map} we show the difference map 
of usual ILC and the input CMB E mode map (for the same seed $100$) weighting each pixel by the 
standard deviation map of global ILC method. The larger pixel values of the bottom panel of
Fig.~\ref{sig_map} compared to the top panel represents a significant  bias present in the reconstructed 
E mode map by the usual ILC method.  

Suppression of chance correlation power in the global ILC method results in accurate estimation
of CMB E mode angular power spectrum. In Fig.~\ref{corr_cl} we present a scatter plot of 
E mode angular power spectra to represent this general feature for only some representative multipoles at different 
ranges as indicated in the figure. For each sub-plot of this figure the horizontal axis represents the 
values of the randomly generated pure CMB E mode angular power spectrum for the specific multipole
indicated in the sub-plot. The vertical axis of a subplot represents the  cleaned power 
spectra values for the corresponding multipole obtained by either global ILC method (golden points)
or by usual ILC method (blue points). The black-line in sub-plot shows the straight line 
with unit slope corresponding maximal correlation between the input and cleaned CMB E mode 
spectrum  for any of the two methods. Two interesting features  of this figure require some emphasis. 
First, at the low multipoles $\ell = 2$ or $3$ the recovered angular spectrum tend to have a value
lower than the input E mode CMB power spectrum for the blue points. This indicates presence of a negative 
bias at the low multipoles in the usual ILC method. The results for the usual ILC method is completely
opposite for the multipoles starting from $\ell \sim 7$ where more and more blue points tend to lie 
above the black line. This implies that  a strong positive bias due to foreground residuals  because 
of presence of the chance-correlations comes into picture in the usual ILC method.  Some 
blue points specifically for multipole starting $\ell = 10$ of this plot lie very far away and above the 
black lines. This further indicates unsatisfactory performance of usual ILC method  to estimate  the E 
mode angular power spectrum at low resolution.   The golden 
points traces the  black-lines closely for all the multipoles  denoting tight correlation between the 
cleaned and the  input  CMB E mode angular power spectra. This indicates a very  desirable property of a CMB reconstruction method 
achieved by the global ILC method of this work, namely, the cleaned angular power spectrum must be   -- 
very strongly correlated with the  pure input CMB E mode spectrum, for an accurate reconstruction of 
CMB E mode angular power spectrum. We show in Fig.~\ref{corr_val}
correlation of cleaned E mode angular power spectra with the input pure CMB spectrum for different
multipoles for the usual ILC and global ILC foreground removal method. The usual ILC spectra becomes 
very weakly correlated with the input CMB E mode spectrum for the entire multipole range $2 \le \ell \le 32$. 
At $\ell = 16$ correlation value for usual ILC drops to merely $32$\%. For the  reconstruction following
global ILC method the minimum correlation is very large $98.4$\%, which occurs for $\ell =20$. The
average correlations for the global method for the entire multipole range is found to be as large as  $99.1$\%.

Each of the two methods -- usual ILC and the global ILC,  produces a set of $1000$ cleaned CMB E mode 
angular power spectra. We estimate two correlation matrices of the cleaned spectra corresponding to these
two sets. The middle and last panel of  Fig~\ref{corr1} shows respectively the correlation 
matrix for the usual ILC and the global ILC method. The figure in the middle panel contains large non-diagonal 
elements (as large as $50$\%) which indicates that the usual ILC cleaned spectra contain strong foreground contamination. The first panel 
of Fig.~\ref{corr1} shows the correlation 
matrix for the input E mode angular power spectra estimated from the $1000$ pure CMB E mode maps. Both the 
first and third panel agree very well with each other. From Fig.~\ref{corr1} we,  therefore, conclude that the cleaned spectra estimated 
from the global ILC method is highly accurate and does not possess any significant foreground residuals.

\section{Discussion and Conclusion}
\label{Con} 

An accurate measurement of CMB E mode polarization signal over large angular scales  can be used to constrain the physics 
of reionization and the decoupling epoch. The CMB E mode signal over the large angular scales 
helps to remove degeneracy between the amplitude of scalar fluctuations and optical depth to 
the reionization epoch. Since the large scale CMB E mode polarization is a weak signal, the problem 
of reconstructing a cleaned E mode CMB map by removing foregrounds  becomes a complex task. Therefore, 
one requires to carefully device an accurate  reconstruction technique so that the cleaned signal
can be reliably used for cosmological analysis. 

In this article, we, for the first time, performed a detailed investigation of reconstruction of CMB E mode
signal over large angular scales  following ILC method by removing astrophysical foregrounds using simulated observations of 
first  $15$ frequency bands of future generation COrE satellite mission. We implemented usual ILC method and an improved and accurate 
global ILC method which uses the prior information of CMB E mode covariance matrix over large 
angular scales of the sky. Our results show that the global ILC method can reconstruct the weak CMB E mode signal accurately over large angular scales  
in contrast to the usual ILC method. 

The empirical  data covariance matrix in the usual ILC method, specifically for CMB E signal reconstruction 
over large angular scales of the sky, contains a chance correlation power between the CMB and non-CMB 
components (e.g., foregrounds). This chance correlation power is very strong at large angular scales over the sky  
due to lower number of available independent modes on these scales. We have shown
in this work, this chance correlation power strongly biases the cleaned E map and  the estimated angular power spectrum 
in the usual ILC method. The chance correlation power merely behaves as random noise and effectively increases rank 
of the empirical  frequency-frequency data covariance matrix leading to large foreground residuals in the 
cleaned E maps. The strong foreground residuals  present in the cleaned E map  implies that the covariance 
matrix of the cleaned map is far from the underlying CMB E mode covariance matrix. In the new global ILC method,
 we overcome  the problem of usual ILC method by incorporating the theoretical CMB E mode covariance matrix in 
 the cleaning algorithm. This causes the cleaned E mode CMB map obtained by using  global ILC method 
to have a  covariance structure  compatible to the underlying theoretical E mode angular  
 power spectrum. Using the underlying covariance 
matrix in our method, therefore, effectively suppresses the  chance correlation power 
and results in the accurate estimation of CMB 
E mode signal and the  full-sky angular power spectrum over the large angular scales.


By doing detailed Monte Carlo simulations we show that there exists a large error in the usual ILC 
cleaned maps due to the CMB-foreground chance correlation power which dominates at large angular scales. 
The error is more than $0.026$ $\mu K$  for almost all regions of the sky. The cleaned power spectrum 
estimated from the cleaned maps obtained following foreground removal using usual ILC method is  biased  
strongly  both due  to a negative bias at the low multipoles and positive bias at higher multipoles.
The error bars on the cleaned E mode power spectrum are in general way larger 
than the cosmic variance prediction for the multipole range $ \ell \ge 7$ of this analysis. The 
new method shows significantly reduced error due to residual foregrounds in the galactic plane. The 
error in the reconstruction of the cleaned E mode map, due to  foreground residuals in the cleaned map,
is somewhat large only in the galactic central region 
where $1\sigma$ error is given by $0.026$ $\mu K$ only at a single pixel . The mean cleaned E mode spectrum from the 
simulations of the global ILC method does not 
show presence of any bias. The $1\sigma$ errors of the E mode angular power spectrum agree very well the 
cosmic-variance induced error alone.   

Not only the cleaned  E maps obtained by using the usual ILC method show presence of significant 
error the angular power spectra estimated from these maps show strong correlation between different 
multipoles indicating presence of foreground residuals in the cleaned angular power spectra. 
A strong positive correlation of about 50\% for either $\ell$ or $\ell'$ larger than $10$ is obtained for the usual ILC case. 
At lower multipoles pair, the effect of negative bias due to CMB-foreground chance correlations leads to 
a weak negative correlation of about $10$\% in the $\ell-\ell'$ matrix. For the cleaned E mode angular power 
spectra obtained using the global ILC method corresponding 
correlation structure is manifest only along 
the diagonal elements of the correlation matrix which agree excellently with the correlation matrix 
obtained from the simulations of pure CMB  E mode angular power spectra. This indicates an accurate 
CMB E mode reconstruction following the global ILC method. 

An interesting statistical feature of the cleaned global ILC  E mode angular power spectra  is that 
the cleaned $C^E_{\ell}$ remains tightly correlated with the corresponding input  pure CMB $C^E_{\ell}$ 
for all multipoles $2 \le \ell \le 32$. This is particularly very advantageous  for the multipole range 
$10 \le \ell \le 25$ where pure E mode $C^E_{\ell}(\ell+1)\ell/2\pi$ is very weak. Presence of strongly outlying
 (erroneous) cleaned angular power spectrum is a general  feature for the usual ILC case. Such outliers are 
entirely absent for the reconstruction performed following global ILC method.

Like the usual ILC method, the global ILC CMB E mode reconstruction does not get affected by the any modelling 
uncertainties in the polarized foreground components.  The global ILC method effectively suppresses the 
CMB-foreground chance correlation power at large angular scales and thereby provides an accurate 
reconstruction of weak CMB E mode map and its angular power spectrum. The method can be extended further 
to obtain a cleaned B mode CMB map and its angular power spectrum. This will be investigated in a future 
publication. 

\section{Acknowledgement}

We  use  publicly  available  HEALPix package \citep{Gorski2005} for the analysis of this work 
(http://healpix.sourceforge.net).  


\label{lastpage}
\end{document}